\documentclass[twocolumn]{aastex62}
\usepackage[utf8]{inputenc}

\usepackage{xcolor}
\usepackage{hyperref}

\received{}
\revised{}
\accepted{}
\submitjournal{ApJS}

\shorttitle{PHOEBE IV: Interstellar extinction}
\shortauthors{Jones et al.}

\begin{document}

\title{Physics of Eclipsing Binaries. IV. The impact of interstellar extinction on the light curves of eclipsing binaries}

\correspondingauthor{David Jones}
\email{djones@iac.es}

\author[0000-0003-3947-5946]{David Jones}
\affiliation{Insituto de Astrof\'isica de Canarias,
E-38205 La Laguna,
Tenerife, 
Spain}
\affiliation{Departamento de Astrof\'isica, 
Universidad de La Laguna,
E-38206 La Laguna,
Tenerife, 
Spain}

\author[0000-0002-5442-8550]{Kyle E. Conroy}
\affiliation{Department of Astrophysics and Planetary Science,
Villanova University,
800 East Lancaster Avenue,
Villanova, 
PA 19085, 
USA}

\author[0000-0002-0504-6003]{Martin Horvat}
\affiliation{University of Ljubljana, Dept. of Physics, Jadranska 19, SI-1000 Ljubljana, Slovenia}

\author[0000-0003-4560-7925]{Joseph Giammarco}
\affiliation{Eastern University, Dept.~of Astronomy and Physics, 1300 Eagle Rd, St.~Davids, PA 19087}

\author[0000-0002-9739-8371]{Angela Kochoska}
\affiliation{Department of Astrophysics and Planetary Science,
Villanova University,
800 East Lancaster Avenue,
Villanova, 
PA 19085, 
USA}

\author[0000-0002-1355-5860]{Herbert Pablo}
\affiliation{American Association of Variable Star Observers, 49 Bay State Road, Cambridge, MA 02138, USA}

\author[0000-0002-3316-7240]{Alex J. Brown}
\affiliation{Department of Physics and Astronomy, University of Sheffield, Sheffield, S3 7RH, UK}

\author[0000-0002-6605-0268]{Paulina Sowicka}
\affiliation{Nicolaus Copernicus Astronomical Center, Bartycka 18, PL-00-716 Warsaw, Poland}

\author[0000-0002-1913-0281]{Andrej Pr\v{s}a}
\affiliation{Department of Astrophysics and Planetary Science,
Villanova University,
800 East Lancaster Avenue,
Villanova, 
PA 19085, 
USA}

\begin{abstract}

Traditionally, the effects of interstellar extinction on binary star light curves have been treated as a uniform reduction in the observed brightness of the system that is independent of orbital phase.  However, unless the orbital plane of the system coincides with the plane of the sky, or if the two stars are completely identical and present with minimal mutual irradiation and tidal/rotational distortions, then this is unlikely to be an accurate representation of the effect of interstellar extinction.  Here, we present an updated treatment of interstellar extinction as incorporated in the PHOEBE 2.2 release (publicly available from \url{http://phoebe-project.org}) and assess the importance of using such an approach in the modeling of different types of binary systems.  We also present the incorporation of PHOENIX model atmospheres into the PHOEBE 2.2 release, providing increased fidelity on computed observables down to lower temperatures than previously available.  The importance of these new code developments is then highlighted via an extincted toy model of the eclipsing white-dwarf-subdwarf binary SDSS~J235524.29+044855.7 -- demonstrating that, in the age of LSST as well as complementary space-based photometric missions, a proper accounting for extinction and as well as the use of realistic model atmospheres will be essential in deriving accurate binary parameters. 
\end{abstract}

\section{Introduction}

Interstellar extinction can have a dramatic effect on the observed flux from an astrophysical source \citep{2003ARA&A..41..241D}, however the studies of binary stars have traditionally applied a rather simplistic approach which, in many cases, will be unable to accurately reproduce observations. The standard approach has been to apply a uniform subtraction to modeled fluxes from a binary star system, irrespective of phase, based on some measure of its extinction (usually based on the object's coordinates). This treatment has been shown to be inadequate, particularly in eclipsing binary stars with large temperature differences between the two components \citep{2005Ap&SS.296..315P}.  A recent example of an improved methodology was used by \citet{maxted18}, where light curve modeling was used to derive the surface brightness ratio between the two stellar components with this then being combined with multi-band photometry and used to estimate the effective temperatures via empirical color-effective temperature and color-surface brightness relations.  This is certainly an improvement on the ``linear'' approach mentioned earlier, but it does rely on fitting priors placed on, for example, limb-darkening -- which may mean that the final modeled parameters may not be entirely consistent with the temperature derived later in the analysis.  

Here we present a revised treatment of extinction as incorporated into the 2.2 release of the PHOEBE code, which allows the model parameters (e.g.\ limb-darkening) to be set in accordance with atmospheric models \citep[i.e.\ allowing the limb-darkening to be set automatically based on the local effective temperature and surface gravity, via interpolated stellar atmosphere models; as described in][]{2016ApJS..227...29P} while also accounting for the impact of extinction on the derived light curve.  Furthermore, we create a selection of toy models to demonstrate the possible impact of extinction on both current and planned survey observations.

\section{Implementation of Extinction in PHOEBE}

The traditional technique to account for interstellar extinction in the modeling of binary star systems is to subtract a constant extinction at all phases, with that value being derived from some estimate of the field extinction (based on the objects coordinates and the preferred extinction map) and the effective wavelength of the filter employed for the observations.  However, the intrinsic spectrum of the observed star system is highly unlikely to be ``flat'' across the wavelength range of the filter, particularly if the filter belongs to a standard broadband system \citep[for example, those employed by the Sloan Digital Sky Survey;][]{2002AJ....123.2121S}.  This non-uniformity introduces a shift in the effective wavelength of the filter based on the shape of the spectrum and, therefore, temperature of the observed star system.  Given that the composite spectrum (i.e.\ including the contributions of both components) of a stellar system can vary significantly with phase (particularly if there is a large temperature differential between the two stars), then the shift in effective wavelength of the filter can also vary with phase.  When one also accounts for the fact that the chosen extinction law can also vary significantly over the filter passband, introducing a further shift in the apparent effective wavelength of the filter in calculating a single extinction value, it is clear that subtracting a constant extinction at all phases is not an adequate solution.  

The only way to accurately account for extinction when modeling a binary star system is to derive the extinction value at each observed phase based on the combination of spectral profile, filter transmission function and reddening law.  PHOEBE derives the local emergent passband intensity for each exposed surface element by multiplying the spectral energy distribution (SED) of that element (based on its effective temperature and either a model atmosphere or blackbody, depending on the user's preference) and the passband transmission function, and finally integrating over wavelength.  This is a computationally expensive process and so, in order to minimize computation time, integrated intensities for a large grid of parameters ($T_\mathrm{eff}$ for blackbodies, and $T_\mathrm{eff}$, log $g$ and [M/H] for model atmospheres) are pre-calculated for a given passband and stored in a look-up table for interpolation at the time of running a binary simulation.

Extinction has been introduced in the 2.2 release of PHOEBE following a similar scheme -- extincted fluxes are precalculated by multiplying the SED by the filter passband and by the extinction law (currently \citealt{1989ApJ...345..245C} for the IR and Optical, and \citealt{2009ApJ...705.1320G} for the UV), and then stored for interpolation at run time.  This is essentially following the methodology employed in PHOEBE since the version 2.0 release for the derivation of passband intensities \citet[outlined in detail in section 5.2.1 of][]{2016ApJS..227...29P}.  The only difference being that the effect of extinction is stored as a multiplicative factor derived as the ratio of integrated passband intensity with and without the application of an extinction law.

In order to account for a wide variety of systems, the extincted fluxes (or rather the multiplicative factor by which passband intensities are altered due to effects of extinction) are derived for the same stellar parameters as for passband intensities (for both blackbodies and model atmospheres) and for a wide range of the visual extinction ($0\leq A_v\leq 10$)  and extinction factor ($2\leq R_v \leq 6$, where $R_v = A_v / E_{B-V}$ and $E_{B-V}$ is the B$-$V color excess).  The user can choose to set value for any two of the three relevant parameters ($R_v$, $A_v$, $E_{B-V}$) in addition to all other parameters of the systems.  If extinction is non-zero, the model passband intensities will be interpolated \cite[again following the scheme outlined in][for the interpolation of passband intensities and limb-darkening, but this time including the additional dimensions defining the extinction, $A_v$ and $R_v$]{2016ApJS..227...29P} from this extended pre-computed grid.

This paper is accompanied by the 2.2 version release of PHOEBE (which is available at \url{http://phoebe-project.org} and on GitHub at \url{https://github.com/phoebe-project}) as well as extinction grids for all currently supported passbands and atmospheres.  In addition to support for extinction, the 2.2 release brings Phoenix model atmospheres \citep{husser13}, numerous improvements and optimizations, as well as added support for both Python 2 and 3.  As support for more passbands and atmospheres are added in the future, we will pre-compute and release the required grids to account for extinction.

\section{Quantifying the impact of extinction}

\subsection{An extreme case}
\label{sec:BK}

In order to highlight the importance of a proper treatment of extinction when simulating binary star light curves, we constructed an array of binary models using PHOEBE 2.2 both with and without any extinction in order to evaluate the difference in eclipse magnitude.  

\begin{figure*}
\centering
\includegraphics[width=\textwidth]{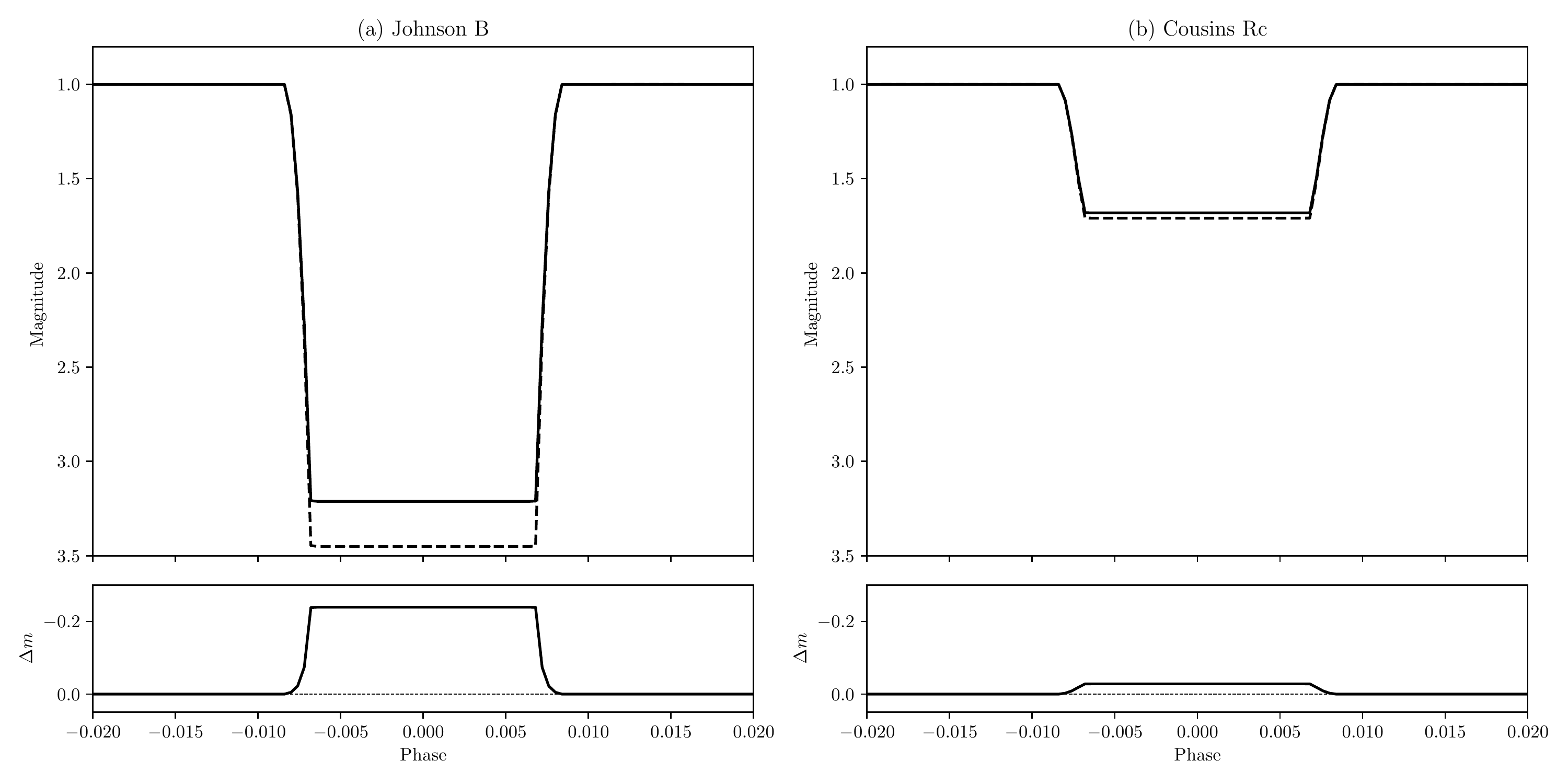}
\caption{Light curve of a synthetic B\,\textsc{v}-K\,\textsc{iii} binary with (solid line) and without (dashed line) extinction correction for $E_{B-V}=1.0$ and $R_v=3.1$ along with residuals between the two cases. Note that out of eclipse magnitudes have been normalised to unity in order to allow a direct comparison of the effect of extinction on the eclipse depth and profile}
\label{fig:BKtest}
\end{figure*}

Let us begin with a rather extreme case, a synthetic binary comprised of a hot, B-type main sequence star ($M=6.5 M_\odot$, $T_\mathrm{eff}=17000$ K and $R=4.2 R_\odot$) and a cool K-type giant ($M=1.8 M_\odot$, $T_\mathrm{eff}=4000$ K and $R=39.5 R_\odot$) in a 1000 day orbit -- a system where, while the temperature difference is large, the luminosities are similar. The synthetic Johnson B and Cousins R light curves for this system - as sampled at 101 evenly spaced intervals between phase -0.02 and 0.02, with the default of 1500 surface triangles for each component (resulting in an execution time per-passband of a few seconds on a single core of a macbook pro) -- both with and without accounting for extinction ($E_{B-V}=1.0$ and $R_v=3.1$), along with the residuals between the two curves, are shown in Figures \ref{fig:BKtest}.   
Clearly, in such a B\,\textsc{v}-K\,\textsc{iii} system, the difference in eclipse depth during primary eclipse (when the K\,\textsc{iii} star passes in front of the B\,\textsc{v}) due to the effect of extinction is appreciable.  In the B-band, the discrepancy reaches more than 0.2 magnitudes -- detectable with even modest observational precision -- while in the R-band the discrepancy is less easily detectable at only $\sim$4 millimag.  In this physically feasible but rather improbable case, limb-darkening was included in the calculation of the light curve in order to later allow direct comparison of the results with synthetic spectra.  

In such a comparison, synthetic spectra of the individual binary components can be used to recreate the spectrum of the system both in and out of eclipse (with and without the effect of extinction) which, combined with the filter transmission profile, can be used to calculate the magnitude of the simulated eclipse.  For both the model PHOEBE light curves and the synthetic comparison spectra, model atmospheres from \citet{2003IAUS..210P.A20C} were used for both components.  For the unextincted system, the eclipse depths derived using PHOEBE match those derived from the synthetic spectra to $\sim$0.1 millimag, confirming that the interpolated limb-darkening function gives results consistent with the disk-integrated synthetic spectra.  

The extincted eclipse fluxes derived using PHOEBE are within a few millimag of those obtained directly from disk-integrated synthetic spectra -- the additional discrepancy arises here from the way that PHOEBE handles the combination of limb-darkening and extinction.  In PHOEBE, each element of the mesh that represents the stellar surface is assigned a flux based on the normal emergent intensity expected from its local thermodynamical and hydrodynamical properties \citep[effective temperature, surface gravity, chemical abundances, etc.][]{2016ApJS..227...29P} with this flux later being adjusted in accordance to a given limb-darkening law -- in the case presented, an interpolation of the comparison of the normal emergent intensity ($\mu=\cos \theta = 1$, where $\theta$ is the angle between the stellar surface normal and the observer) and the specific intensity at a variety of positions along the stellar limb ($0>\mu>1$).  This interpolated limb-darkening means the results are comparable to disk-integrated synthetic spectra \citep{1994AJ....107..742G}, which are the result of integrating the specific intensities across the stellar surface (i.e. with respect to $\mu$).  

In the case of an extincted system, however, PHOEBE applies an extinction factor calculated from the user-provided extinction parameters ($A_V$ and $R_v$) to each surface element based on those tabulated for normal emergent intensities ($\mu=1$) which are then applied along with the limb-darkening factors. Collectively, this means that towards the stellar limb ($\mu\to0$), the applied extinction factors begin to deviate from the values that would be determined from a synthetic spectrum of that point on the stellar surface (as the shape of the spectrum varies slightly as a  function of $\mu$).  The effect of this deviation, however, is clearly very small (less than 1\%) and certainly smaller than the uncertainty on any measured extinction ($A_V$) used as an input to the model as well as on the generalised extinction law employed \citep{1989ApJ...345..245C,2009ApJ...705.1320G}.  As such, we conclude that the current application of extinction introduced here in the 2.2 release of PHOEBE is valid and consistent with expected results.

\begin{figure}
\centering
\includegraphics[width=0.5\textwidth]{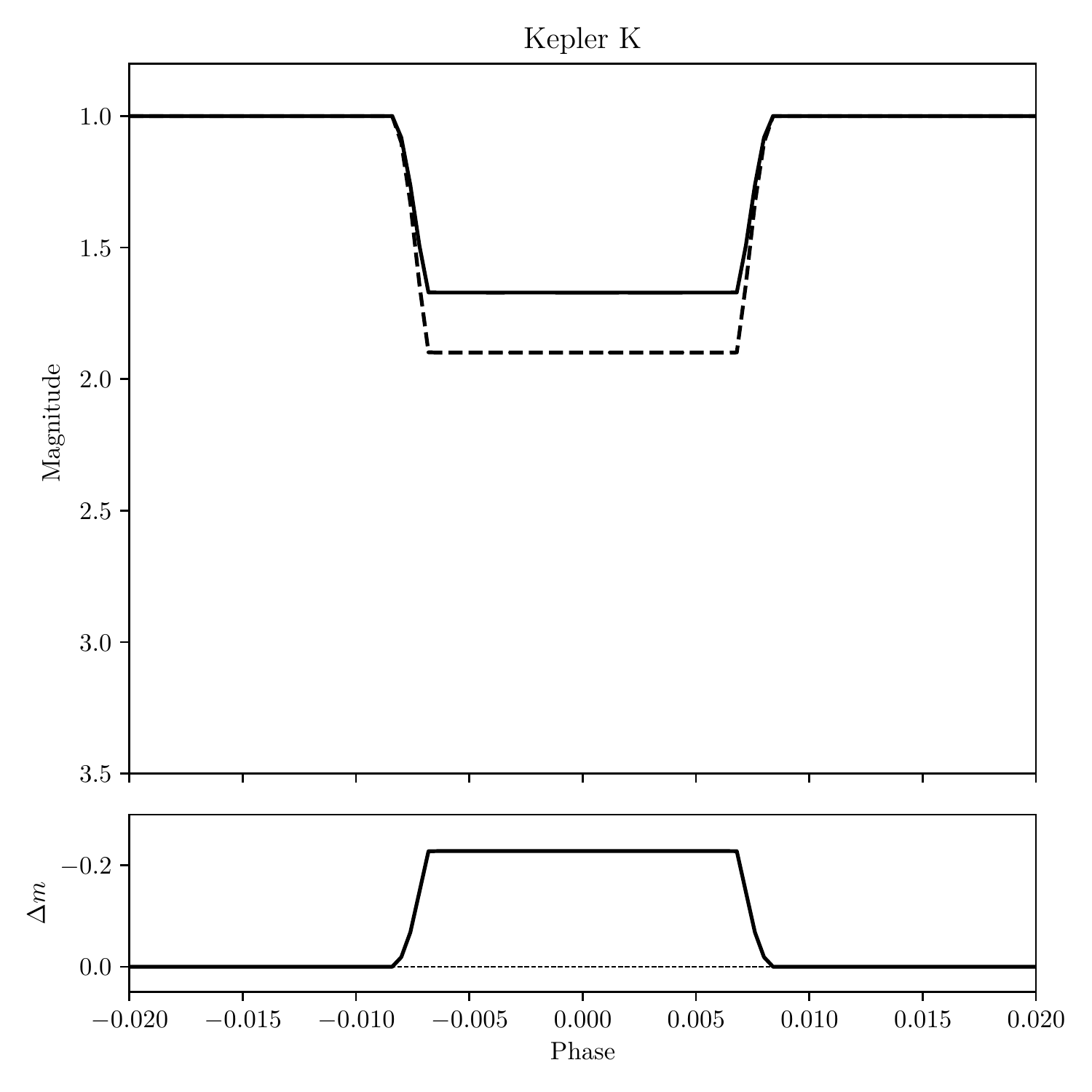}
\caption{Light curve of synthetic B\,\textsc{v}-K\,\textsc{iii} binary with (solid line) and without (dashed line) extinction correction for $E_{B-V}=1.0$ and $R_v=3.1$ along with residuals between the two cases.}
\label{fig:BK_kep}
\end{figure}

Given the marked difference between the B- and R-bands, one may also wish to consider the impact in a much broader passband -- that of the Kepler satellite being a good example (covering more or less the same wavelength range as the Johnson B-,V- and R-bands).  The same B\,\textsc{v}-K\,\textsc{iii} eclipse as would be observed using with the Kepler passband is shown in figure \ref{fig:BK_kep}.  Intriguingly, in spite of the much broader wavelength coverage, the difference between extincted and unextincted light curves is still a few tenths of a magnitude -- easily detectable at normal photometric precision of Kepler.

\subsection{``Normal'' binary systems}
\label{sec:solar}

\begin{figure*}
\centering
\plottwo{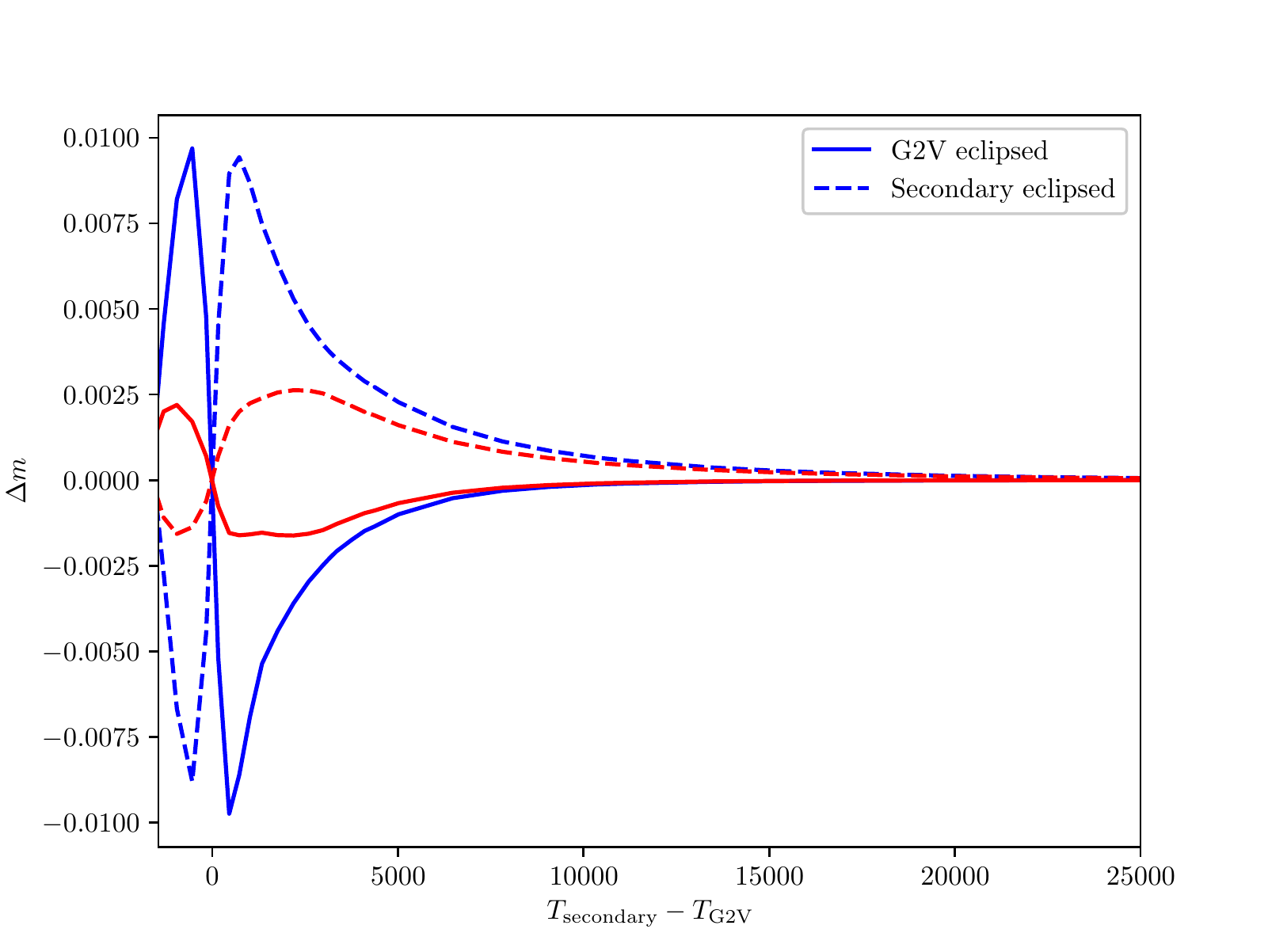}{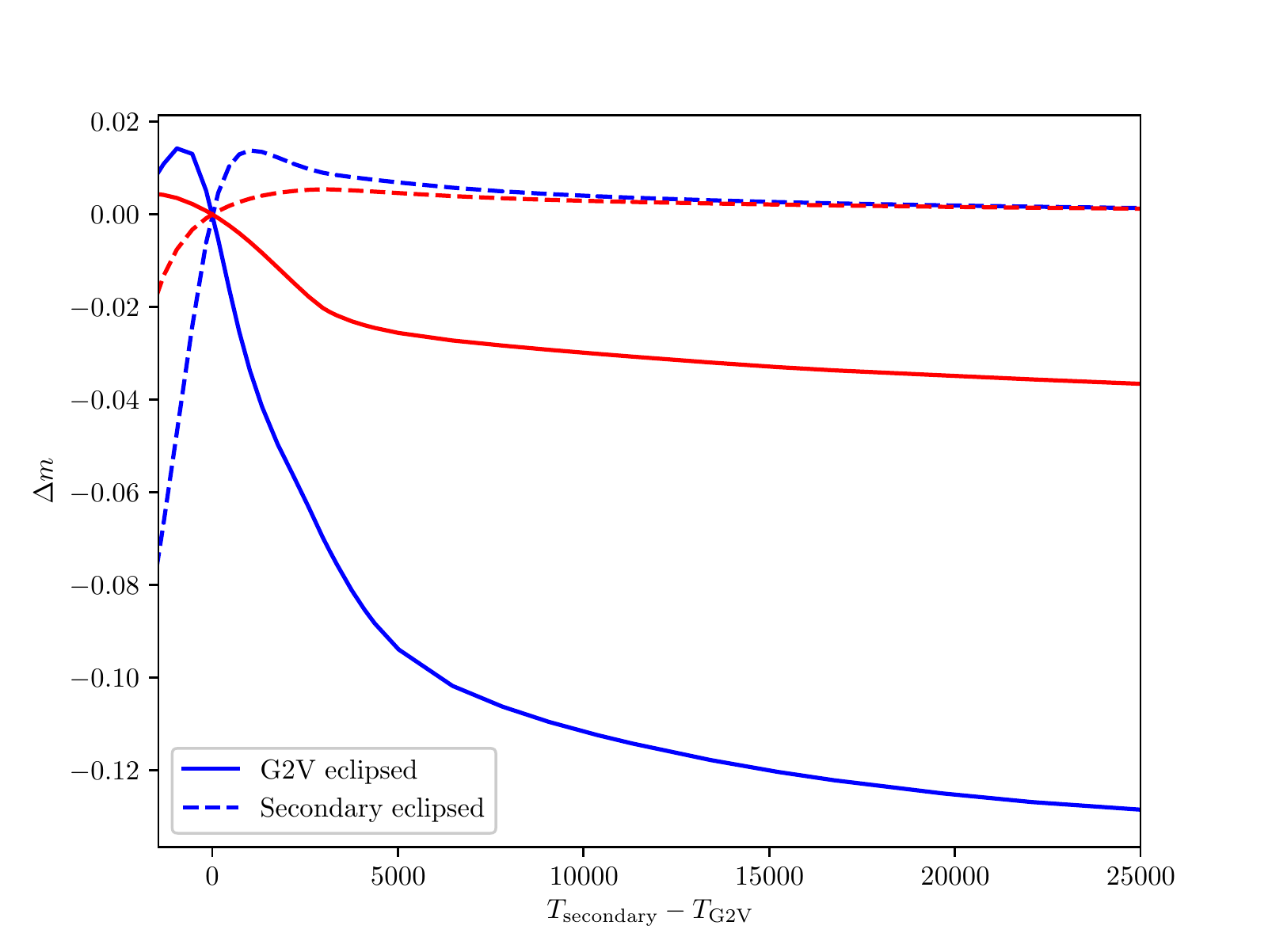}
\caption{The difference in eclipse magnitude (both primary and secondary) when an extinction of $E_{B-V}=1.0$ is accounted for or not as a function of temperature difference between the two binary components.  The blue curves show the results for a Bessell B filter, while the red curves show the results for a Bessell R filter.  The synthetic binaries used to calculate the curves on the left plot comprised a Sun-like star (always considered to be the primary) and a zero age main sequence companion \citep{2008A&A...484..815B,2009A&A...508..355B}, while those on the right comprised a Sun-like star (again always considered the primary) and an identical (though not entirely physical) companion where only the effective temperature was changed.}
\label{fig:G2Vtest}
\end{figure*}

Returning to the importance of accounting for extinction in the modeling of binary light curves: while the previously highlighted B\,\textsc{v}-K\,\textsc{iii} case is a rather extreme one it does demonstrate that extinction can have an appreciable effect on the observed light curve of an eclipsing binary. Of course, not all systems will present such extreme discrepancies, so let us consider some more ``normal'' binary systems comprising a solar-type star (G2V, $T_\mathrm{eff}=5780$ K, Mass$=1 M_\odot$, Radius$= 1 R_\odot$) and a main sequence companion \citep[with parameters taken from the zero age main sequence of][]{2008A&A...484..815B,2009A&A...508..355B} in an eclipsing ($i=90$\degr{}) 10 day orbit.  All other second-order effects were disabled in the evaluation of the light curves (limb-darkening, gravity-brightening, irradiation effects and tidal distortions), such that the effect of extinction could be isolated.  The measured differences in the eclipse depth for zero extinction and an $E_{B-V}=1.0$ ($R_v=3.1$) of the solar-type star and its companion in both Johnson B and Cousins R filters are shown in figure \ref{fig:G2Vtest} (left).

As expected, the impact of extinction is more evident in the B-band than R-band as a result of the strong wavelength dependence of the employed extinction law \citep{1989ApJ...345..245C,2009ApJ...705.1320G}.  However, the synthetic light curves also show that, for main sequence binaries, the effect does not scale with the temperature difference between the two components.  This may seem counter-intuitive, but is actually a reflection of the difference in luminosity and radius with effective temperature along the main sequence. For example, consider a much hotter companion of roughly three times the temperature of the G2V star \citep[corresponding to an early B-type main sequence star of roughly 5 M$_\odot$ on the zero age main sequence of ][]{2009A&A...508..355B}.  While the temperature difference does result in different extinction corrections for the two components, this is almost completely negated by the combination of increased radius (nearly 3 R$_\odot$, meaning that even when eclipsed, much of the early B-type companion is visible) and luminosity (roughly 500 L$_\odot$) which act to ensure that the companion dominates the observed flux at all phases.  

To isolate the dependence on effective temperature, a series of binary models were constructed with the same components as before but this time fixing the mass and radii of both stars to be solar (i.e. only varying the effective temperature of the companion).  The majority of these systems are clearly non-physical, but serve to highlight the dependence on effective temperature difference with the results plotted in figure \ref{fig:G2Vtest} (right).  Overall, the curves show that the effect scales non-linearly with temperature difference, increasing rapidly in magnitude for similar temperatures but beginning to plateau for large temperature differences.  This is an effect of the variation in the shape of the stars' SEDs in the B- and R-bands as a function of temperature -- once the temperature of the secondary becomes large, the shape of its SED in these optical bands begins to stabilize, with the largest variations seen around the peak of the SED (now deep in the ultraviolet).  This effect is also shown by the ``peaks'' in the impact of extinction found at relatively small temperature differences -- these reflect the fact that around these points, the peak of the companion's SED is within the filter passband (most obvious in the B-band given that the primary's SED also peaks in this band).

While, as mentioned previously, the models with non-varying secondary mass and radius are clearly non-physical, when combined with the previous (more physically sound) models, they do serve to reinforce that the effect of extinction can be appreciable.  In extreme cases, like the hot main sequence -- cool giant system presented here or pre-cataclysmic variables which comprise relatively large subdwarfs with low mass main sequence companions, the discrepancy between extincted and unextincted eclipse depths can reach a few tenths of a magnitude or more (detectable with even modest instrumentation).  For more normal, main sequence binaries, the effect is smaller being of order 1--10 millimagnitudes for an $E_{B-V}=1$.  These cases are slightly more challenging to detect, but still well within the capabilities of modern facilities \citep[e.g.\ GAIA;][]{2017A&A...599A..32V}.

Additionally, the models presented highlight that the effect of extinction is highly non-linear and, as such, extremely difficult to account for after the fact and, therefore, should ideally be incorporated into the modeling effort from the beginning.

\subsection{Large Synoptic Survey Telescope / Vera C. Rubin Observatory}
\label{sec:LSST}

It should already be clear that the effect of extinction will be appreciable in very high-precision space-based photometry (see e.g., the final paragraph of Section \ref{sec:BK}).  However, in most cases \citep[e.g.\ Kepler, TESS;][]{kepler,tess} these observations are delivered only in a single passband, often meaning that -- while extinction must be accounted for if one is to derive precise binary parameters \citep[see e.g.;][]{armstrong14,maxted18} -- a model can still be fit to the data even without accounting for extinction (although the model parameters, particularly temperatures, will almost certainly not be accurately derived due to the impact of extinction).  Multi-band surveys, even from the ground, are perhaps more likely to show discrepancies which make modeling impossible without accounting for extinction (as fits to individual bands, ignoring extinction, will almost certainly require different model temperatures).

The Large Synoptic Survey Telescope (LSST) or Vera C.\ Rubin Observatory is an in-construction facility which aims to survey the entire sky in six photometric bands with a three-day cadence.  The expected precision for the $g$,$r$ and $i$ bands is expected to be roughly 5 millimag, and approximately 7.5 millimag in the $u$, $z$ and $y$ bands \citep{iveziclsst}.  Given the shape of the extinction law, the observed extinction effect in the $i$, $z$ and $y$ bands is likely to be minimal in all but the most extreme cases.  However, in the $u$, $g$ and $r$ bands even some main sequence binaries, such as those presented in Section \ref{sec:solar}, could be impacted by extinction.  To highlight this, we construct a selection of ``normal'' realistic synthetic binaries based on the orbital periods, masses, radii and effective temperatures of the systems found in John Southworth's catalogue of detached eclipsing binaries DEBCat \citep{southworth15}\footnote{Due to the inhomogeneous nature of the data in the catalogue, we ignore metallicity and assume zero eccentricity and inclinations of 90$^\circ$.}.  We then compare the synthetic light curves for this system in the LSST bands for zero extinction and for an extinction consistent with the Galactic bulge \citep[$A_V=2$, $R_v$=2.5;][]{sumi04,OGLEext}.

As one might predict based on the curves presented in Figure \ref{fig:G2Vtest}, the majority of systems show discrepancies due to extinction on the order of 1 millimag -- below the LSST detection limit.  However, some systems would present detectable differences due to extinction.  For example, the main sequence binary ZZ UMa, which comprises an F8 primary and G6 secondary \citep{ZZUma}, would present with a borderline detectable discrepancy in the g-band where the primary eclipse would be observed to be $\sim$7 millimag shallower and the secondary eclipse $\sim$6 millimag deeper.  A more pronounced effect would be found in the sub-giant-main-sequence binary AI Phe \citep{AIPhe}, where the g-band secondary eclipse would be observed to be $\sim$19 millimag shallower (a more than 3$\sigma$ difference) and the primary eclipse $\sim$6 millimag deeper.  

In all cases, the discrepancies in u-, r-, i-, y- and z-band eclipse depths were found to be lower than the LSST detection limits.  This is not unexpected for the redder filters (r-, i-, y- and z-bands), where extinction is minimal, but is perhaps surprising in the u-band.  However, this is a consequence of both the poorer precision expected in this band (7.5 millimag) and the fact that DEBCat does not contain any binaries with early-type primaries and late-type companions (i.e.\ the systems where the greatest change in the u-band SED would be observed during an eclipse).  

\begin{figure*}
\centering
\includegraphics[width=\textwidth]{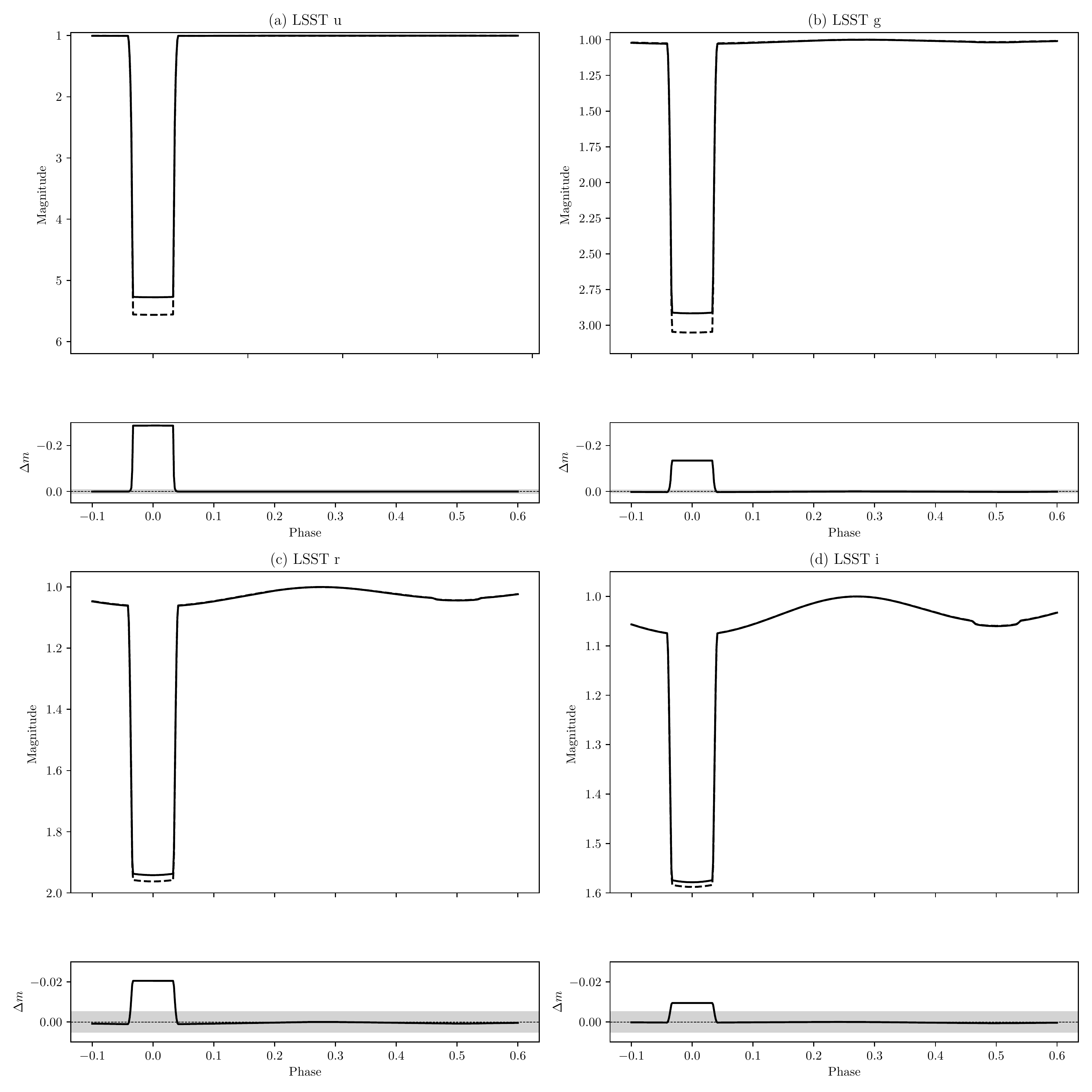}
\caption{Synthetic LSST light curves of the white-dwarf-subdwarf binary SDSS~J2355 \citep{rebassa-mansergas19}, calculated without (dashed line) and with extinction (solid line) corresponding approximately to the average extinction of the Galactic bulge.  The residuals between extincted and unextincted cases are shown below each light curve, highlighting that the difference would be detectable at several $\sigma$ by LSST in u, g and r-bands, and marginally detected in the i-band (in each residual plot, the shaded region represents $\pm 1 \sigma$).} 
\label{fig:rebassa}
\end{figure*}

As such, it is perhaps more interesting to consider some more systems where the components are in different evolutionary phases -- for example, a white-dwarf-main-sequence (WDMS) system.  As highlighted by \citet{gianninas13}, LSST could expect to find of order 100\,000 white dwarfs with stellar mass companions -- meaning that such systems are far from extraordinary.  Furthermore, many WDMS systems will be post-common-envelope systems \citep[perhaps still even being surrounded by the remnant common envelope in the form of a planetary nebula;][]{jones17,boffin19} which may lead to increased local extinction as a result of dust formation in the ejected envelope \citep{lu13}.  As such, they are rather appropriate systems to constrain the importance of extinction in deriving their stellar parameters.

An important caveat that must be noted is that many of these WDMS systems could also exhibit significant levels of irradiation -- a problematic effect when it comes to the treatment of extinction as irradiation alters the emergent stellar spectrum, while here we only consider non-irradiated atmosphere models \cite[see the discussion of][for more details]{horvat19}\footnote{As the discussion here is considered only illustrative, we fix the bolometric albedos and bolometric limb-darkening coefficients of both components to standard values.}.

As a starting point, let us take one of the best constrained eclipsing binaries known -- SDSS~J235524.29+044855.7 \citep[hereinafter referred to as SDSS~J2355;][]{rebassa-mansergas19}.  SDSS~J2355 is a short-period post-CE binary comprising a relatively cool white dwarf ($T_\mathrm{eff}\sim13,250$ K) and a low-mass, metal-poor, sub-dwarf star (spectral type $\sim$sdK7).  As before, calculating synthetic light curves for the system with no extinction and then with extinction consistent with the Galactic bulge, we now see significant deviations between the two models in u, g and r bands (See figure \ref{fig:rebassa}).  The u-band shows the strongest difference at more than 0.25 magnitudes during primary eclipse, closely followed by the g-band (at 0.13 magnitudes) and the r-band (at 0.02 magnitudes, approximately 4$\sigma$).  Even in the i-band, the deviation is almost 0.01 magnitudes -- representing a more than 1$\sigma$ difference.  The deviation during secondary eclipse would not be detected in any band given that the eclipse is not complete and the primary is far more luminous.

\section{PHOENIX atmospheres}

To increase the fidelity of computed observables on the low-temperature end, PHOEBE 2.2 now incorporates PHOENIX model atmospheres \citep{hauschildt1997} as computed by \citet{husser13}. PHOENIX is one of the leading alternatives to \citeauthor{2003IAUS..210P.A20C} that incorporates 3D NLTE model atmospheres, and has already been employed for binary light curve synthesis by, for example, \citet{orosz2000}. The models themselves span wavelengths between $500${\AA} and $5.5\mu$m, effective temperatures between 2300\,K and 12,000\,K, surface gravities ($\log g$) between 0.0 and 6.0, and metallicities ([Fe/H]) between -4.0 and 1.0.

\begin{figure*}
    \centering
    \includegraphics[width=\textwidth]{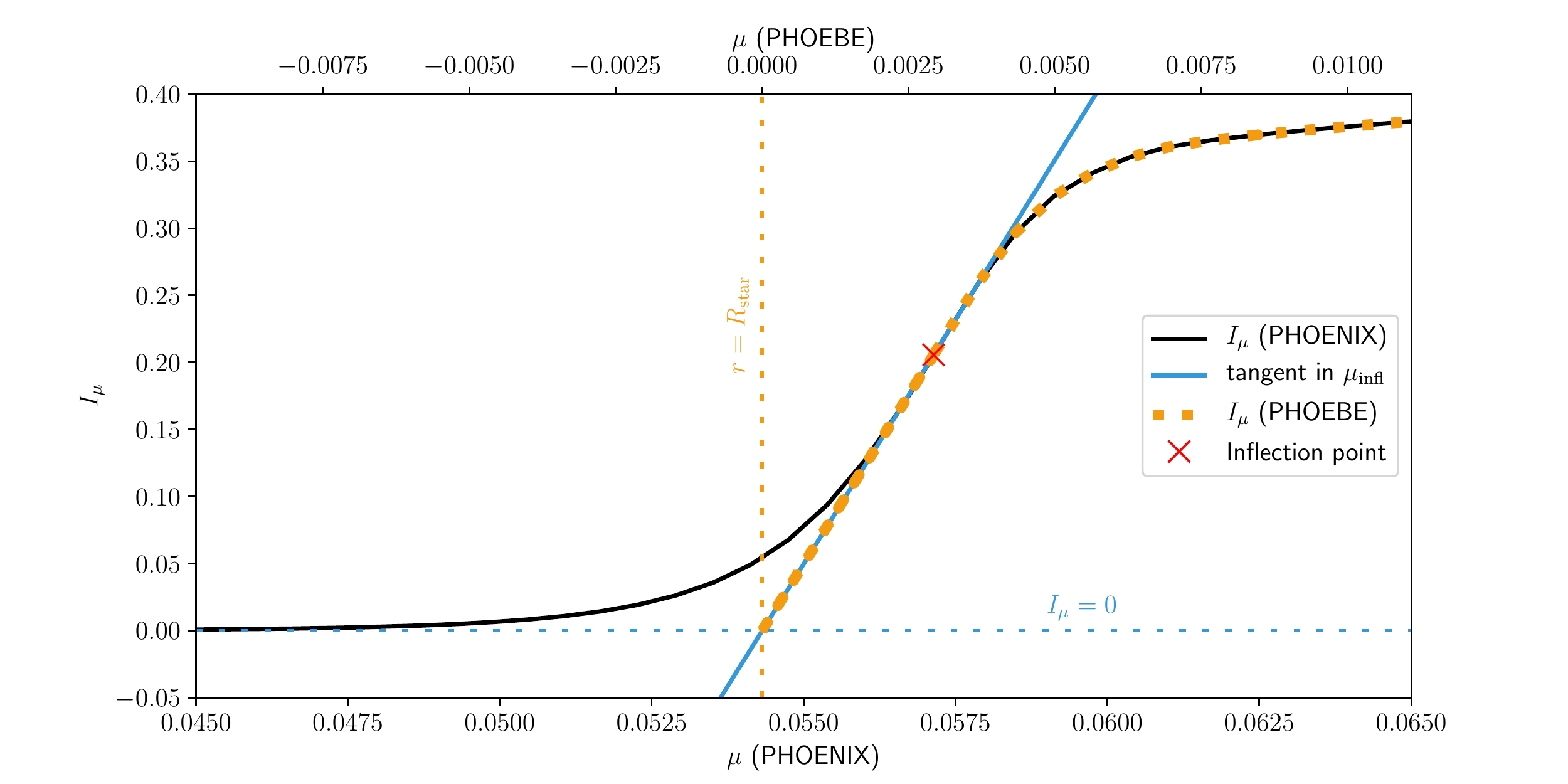}
    \caption{Rescaling of PHOENIX emergent directions (bottom $x$-axis) to PHOEBE emergent directions $\mu$ (top $x$-axis). We find the inflexion point of the PHOENIX $I_\mu$ and calculate a tangent in that point. PHOEBE's $\mu=0$ is then set at the intersection of the tangent and the $x$-axis.}
    \label{fig:tangent}
\end{figure*}

Two challenges presented themselves for the incorporation of PHOENIX model atmospheres. First, the computational scheme for PHOENIX 3D model atmospheres defines emergent direction $\mu$ with respect to the computational grid rather than the stellar mesh, so emergent specific intensity $I(\mu)$ tends to 0 \emph{before} $\mu \to 0$. This is done because the size of the star depends on the wavelength and the computational grid is chosen to accommodate the largest radii. However, this poses a problem because PHOEBE treats stellar surfaces as opaque, and $\mu$ at the limb is defined as 0. Thus, PHOENIX values of $\mu$ had to be rescaled to PHOEBE values of $\mu$ in a way that would ``shrink'' the computational grid to the radius of the star. To achieve this, we calculate a tangent in the inflection point of $I(\mu)$ and set $\mu \equiv 0$ at the intersection of that tangent with the $I = 0$ axis (cf.~Fig.~\ref{fig:tangent}). That workaround imposes $I(\mu=0)=0$ that PHOEBE requires for its operation. Any truncated contribution to intensity leftwards of the intersection point is below a small fraction of a percent.

Second, model atmospheres for some combinations of atmospheric parameters ($T_\mathrm{eff}$, $\log g$ and metallicity) in the \citeauthor{husser13} table were not available at the time of this writing. To overcome that, we compute the look-up tables (as discussed in Section 2) and then \emph{impute} the missing values by triangulating the available data using a convex hull and performing linear barycentric interpolation to estimate the missing values.

PHOENIX atmospheres are now fully available for the computation of observables in PHOEBE. Much like \citeauthor{2003IAUS..210P.A20C}, we make the computations available in both energy-weighted and photon-weighted regime (see \citet{2016ApJS..227...29P} for details), and we incorporate full support for extinction following the scheme previously outlined. To use PHOENIX model atmospheres instead of \citeauthor{2003IAUS..210P.A20C}, \texttt{atm=phoenix} should be passed to PHOEBE instead of \texttt{atm=ck2004}.

\section{Summary}

We have presented the implementation of interstellar extinction as incorporated into the PHOEBE 2.2 release.  Through the construction of synthetic binary systems, and comparison with calculations based on integrated model atmospheres, we have demonstrated that the current implementation is self-consistent to better than 1\%.  We then go on to highlight some cases in which accounting for interstellar extinction could be critical in deriving accurate stellar parameters (effective temperatures, in particular).  Considering moderate extinctions consistent with the average extinction of the Galactic bulge, we show that for most main sequence binaries the difference between extincted and unextincted light curves is relatively small and likely only detectable from space (or borderline detected in a single-band with LSST).  However, when the two components of the binary are in different evolutionary phases, the difference between extincted and unextincted curves can be appreciable.  For example, the subgiant-main-sequence binary AI Phe would show detectable differences between extincted and unextincted cases (at approximately 3$\sigma$ in the LSST g-band).  For more extreme evolutionary phase differences, like WDMS binaries -- some 100\,000 of which are expected to be observed by LSST, the effect of extinction should be appreciable at even greater levels in multiple bands (principally u, g and r-band), emphasising the importance of a proper accounting of extinction in the analysis and modeling of such ground-based survey data as well as space-based photometry.

We similarly present the incorporation, in PHOEBE 2.2, of a grid of model atmospheres based on the PHOENIX models \citep{hauschildt1997}.  These atmospheres represent one of the leading alternatives to the \citet{2003IAUS..210P.A20C} atmospheres already available, in particular extending the range of usable effective temperatures downwards to 2300~K (c.f.\ 3500~K from \citeauthor{2003IAUS..210P.A20C}).  The combined importance of this extended temperature range over which model atmospheres can be used, as well as the treatment of extinction already outlined, is highlighted using an extincted toy model of one of the best constrained eclipsing binaries known, the white-dwarf-subdwarf binary SDSS~J235524.29+044855.7 \citep{rebassa-mansergas19}.  Ultimately, the code developments presented extend the range of stellar and systemic parameters over which PHOEBE can provide increased model fidelity.

\acknowledgments

The development of PHOEBE is possible through the NSF AAG grants \#1517474 and \#1909109 and NASA 17-ADAP17-68, which we gratefully acknowledge.  DJ acknowledges support from the State Research Agency (AEI) of the Spanish Ministry of Science, Innovation and Universities (MCIU) and the European Regional Development Fund (FEDER) under grant AYA2017-83383-P.  DJ also acknowledges support under grant P/308614 financed by funds transferred from the Spanish Ministry of Science, Innovation and Universities, charged to the General State Budgets and with funds transferred from the General Budgets of the Autonomous Community of the Canary Islands by the Ministry of Economy, Industry, Trade and Knowledge. PS thanks the Polish National Center for Science (NCN) for support through grant
2015/18/A/ST9/00578.
The authors thankfully acknowledge the technical expertise and assistance provided by the Spanish Supercomputing Network (Red Espa\~nola de Supercomputaci\'on), as well as the computer resources used: the LaPalma Supercomputer, located at the Instituto de Astrof\'isica de Canarias.

\vspace{5mm}

\software{  PHOEBE \citep{2016ApJS..227...29P,horvat19},
            SPECTRUM \citep{1994AJ....107..742G},
            astropy \citep{2013A&A...558A..33A},
            matplotlib \citep{2007CSE.....9...90H},
            numpy   \citep{2011CSE....13...22},
          }

\bibliography{refs}

\end{document}